# Background Stratospheric Aerosol Investigations Using Multi-Color Wide-Field Measurements of the Twilight Sky

Oleg S. Ugolnikov and Igor A. Maslov

Space Research Institute, Russian Academy of Sciences
Profsoyuznaya st., 84/32, Moscow 117997 Russia
E-mail: ougolnikov@gmail.com

First results of multi-wavelength measurements of the twilight sky background using all-sky camera with RGB-color CCD conducted in spring and summer of 2016 in central Russia (55.2°N, 37.5°E) are discussed. They show the effect of aerosol scattering at altitudes up to 35 km which significantly increases to the long-wave range (624 nm, R channel). Analysis of sky color behavior during the light period of twilight with account of ozone Chappuis absorption allows retrieving the angle dependencies of scattering on the stratospheric aerosol particles. This is used to find the parameters of lognormal size distribution: median radius about 0.08 microns and width 1.5-1.6 for stratospheric altitude range.

**1. Introduction**

It is well-known that most part of aerosol particles in the atmosphere of Earth is distributed in its lower layer, the troposphere. However, upper atmospheric layers are not absolutely free from solid or liquid particles. As early as in late XIX century, after Krakatoa eruption in 1883, the color change of the twilight sky was noticed (Clark, 1883), the phenomenon was called "volcanic purple light" (Lee and Hernádez-Andrés, 2003). Gruner and Kleinert (1927) explained it by aerosol light scattering above the troposphere.

Existence of aerosol layer in the lower stratosphere was confirmed in balloon experiments (Junge et al, 1961), and it was called Junge layer. Aerosol was detected at different latitudes and thus could not be contained of pure water or ice, since the temperature is above the water condensation threshold (except polar regions in winter). As it was shown by Rosen (1971), it is the solution of sulfuric acid. It is produced by the chemical reactions of sulfur dioxide transferred to the stratosphere from the ground.

Long series of balloon observations (Deshler et al., 2003) had shown that the particle size distribution was monomodal with median radius about 0.1 μm. Major eruptions in late XX century (El Chichon in 1982, Mt. Pinatubo in 1991) rapidly changed the physical and optical characteristics of stratospheric aerosol (Jager, Deshler, 2002; Deshler et al., 2003; Bauman et al., 2003). The particle size distribution became bimodal with larger fraction size more than 0.3 μm, maximal size exceeded 1 μm. Large amount of aerosol scattered the solar emission, decreasing the Earth's surface temperature (Hansen et al., 1992). It also destructed the stratospheric ozone by means of heterogeneous chemical reactions (Hoffman and Solomon, 1989).

One of possible sources of background aerosol particles is anthropogenic sulfur dioxide (Brock et al., 1995). This relates the question of stratospheric aerosol with global climate change. Hoffman and Rosen (1980) noticed the possible increase of background aerosol compared with early observations (Junge et al., 1961). The gradual increase was also observed recently, during the beginning of XXI century (Solomon et al., 2011).

Background stratospheric aerosol particles also play important role in physics and chemistry of the middle atmosphere. Reflection of solar radiation is basically related with small particles those scatter significant part of light to the back hemisphere (Hinds, 1999). In high latitudes tiny particles also play the role of condensation nuclei for polar stratospheric (or nacreous) clouds strongly influencing the ozone chemistry.

The backscatter ratio of total and Rayleigh scattering measured in lidar experiments (Zuev et al., 2007, Burlakov et al., 2011) is not more than 1.2-1.3 after moderate eruptions like Tavurvur in 2006 and other recent events. During the volcanically quiet epochs, stratospheric aerosol backscattering is quite small admixture to the Rayleigh level. However, it increases at lower scattering angles owing to properties of Mie scattering. Wavelength dependence of the intensity differs from the one of Rayleigh scattering and can be also used for size estimation, that was performed in limb measurements (Thomason et al., 2007; Bourassa et al., 2008) and lidar sounding (Von Zahn et al., 2000, Jumelet, 2008).

Background aerosol scattering also changes the characteristics of the twilight sky: brightness distribution, color, and polarization. This can be used to separate this component of the twilight background and to find its observational characteristics (Ugolnikov and Maslov, 2009). In that paper the polarization of stratospheric aerosol scattering was estimated after the Tavurvur (Rabaul) volcano eruption in 2006 (0.28±0.03 for scattering angle 92° and wavelength 525 nm). Given the type of size distribution of aerosol particles (lognormal with $\sigma=1.6$), the mean radius value (0.107±0.005 microns) can be estimated with high accuracy.

Twilight geometry of radiation transfer allows separating the definite atmospheric level still illuminated by the Sun while lower dense layers are immersed into the Earth's shadow. Use of all-sky cameras helps to cover wide range of scattering angles, this is important for particle size estimation. It gets more exact if multi-color data are also used. Multi-wavelength observations are of special interest not only because of wavelength dependence of aerosol scattering. Strong excess of Rayleigh scattering in shorter wavelengths leads to the significant difference of effective scattering altitude in blue and red bands. The same atmospheric volume can be still illuminated by straight solar radiation in red band and be strongly obscured in blue band. Blue spectral region is also characterized by higher contribution of multiple scattering (Ugolnikov, 1999; Ugolnikov and Maslov, 2002).

The effects listed above are the reasons of intensive red color of thin clouds at sunset. The stratospheric aerosol particles can influence the sky color during the deeper stage of twilight, at solar zenith angles about 92-93°. It can be detected as color change in the dawn segment. However, color of the sky is also influenced by Chappuis absorption bands of atmospheric ozone and effects of multiple scattering those must be also taken into account. The basic aim of this paper is to fix the size distribution of stratospheric aerosol particles basing on simple color measurements of the twilight sky.

**2. Observational effect of stratospheric aerosol**

Measurements of the twilight sky background are being conducted in Chepelevo, central Russia (55.2°N, 37.5°E) using all-sky camera described by Ugolnikov and Maslov (2013ab). This camera is designed for measurements over the wide part of the sky with diameter about 140°. During the spring and early summer of 2016, RGB-color Sony QHYCCD-8L matrix was installed. Effective wavelength of B, G and R channels was equal to 461, 540 and 624 nm, respectively (R-channel is also corrected by IR-blocking filter). The diameter of sky image is about 650 pixels. Exposure time varied from 3 ms at sunset/sunrise to 30 s during the night. Camera position, flat field and atmospheric transparency are controlled by star images photometry in the night frames.

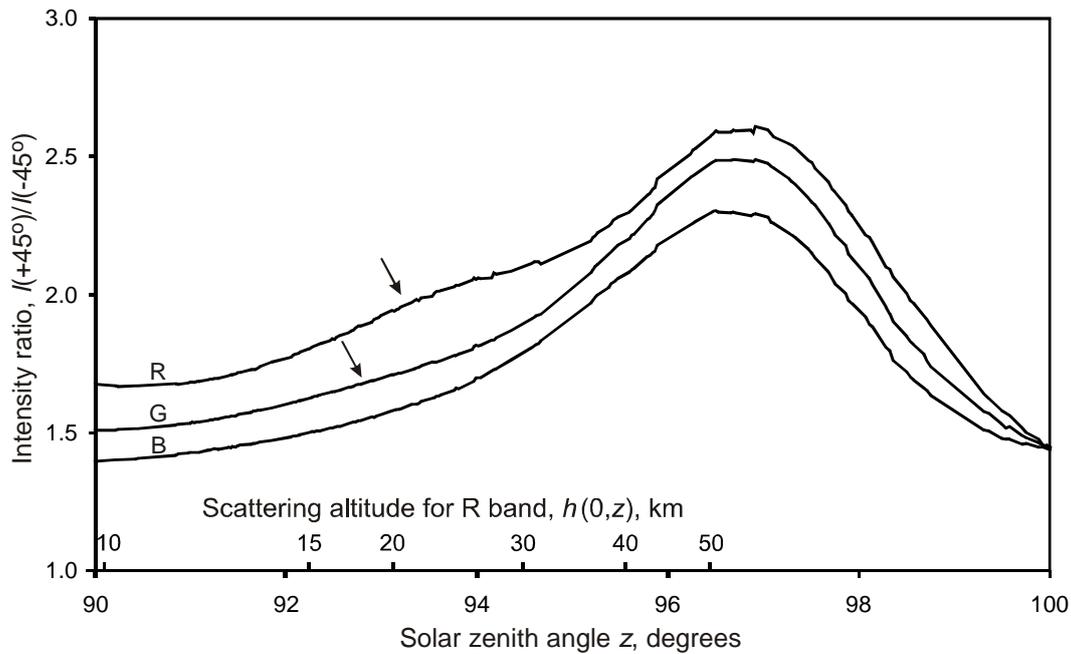

*Figure 1. Twilight sky brightness ratio in symmetric points of solar vertical (evening twilight of March, 27, 2016). Arrows show the effect of aerosol scattering in stratosphere.*

This paper is based on the results of measurements in the solar vertical. In this case the sky point position is characterized by the coordinate $\zeta$ (Ugolnikov, Maslov, 2013b), equal to zero in the zenith, positive in the dusk/dawn area and negative in the opposite sky part. The module of this value is equal to the zenith distance of the observational point in the solar vertical.

Figure 1 shows the dependence of sky brightness ratio in the symmetric points of the solar vertical $I(\zeta=+45°)/I(\zeta=-45°)$ for all three channels during the evening twilight of March, 27, 2016. This dependence was described in (Ugolnikov, 1999) and reflects the behavior of ratio of single and multiple scattering intensities. During the light stage of twilight, the ratio $I(+\zeta)/I(-\zeta)$ increases due to appearance of difference of single scattering altitudes in the dusk area and opposite sky part. During the darker stage, at solar zenith angles more than 96°, the brightness excess in the dusk area decreases as the single scattering fades on the background of multiple scattering.

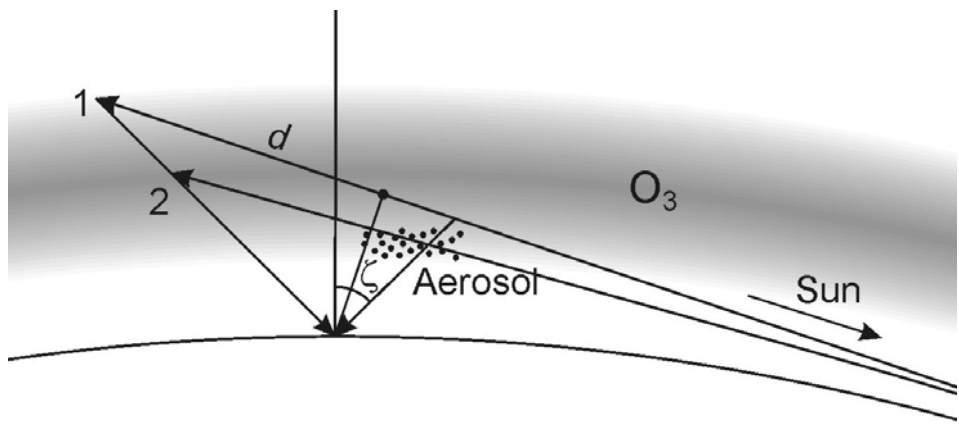

*Figure 2. On the explanation of color effects of the sky background during the twilight.*

However, this dependence for R channel has a remarkable feature at solar zenith angle about 93°: additional brightness excess in the dusk area (arrows in Figure 1) that is barely visible in G channel and fades in B channel. During this twilight stage, effective scattering takes place at the altitudes about 20 km, close to Junge aerosol layer. Twilight sky color is also strongly influenced by Chappuis bands of stratospheric ozone. However, it has the same order of value in G band, but the dusk excess of brightness is significantly less there. We can see that this excess is shifted to the lower values of solar zenith angle in G channel. This is related with higher position of effective scattering layer compared with R channel. So, the same aerosol layer corresponds to earlier twilight stage in G channel.

Figure 2 presents the graphical scheme of single scattering during the light stage of twilight. Before being scattered, the solar emission goes through the stratosphere almost horizontally. The perigee height of effective path decreases with wavelength owing to sharp spectral dependence of Rayleigh scattering. The length of path fraction through ozone layer is long (line 1 in Figure 2), Chappuis absorption in green and red spectral range and multiple scattering effects (Ugolnikov et al., 2004) lead to gradual "bluing" of sky spectrum from dusk/dawn area to opposite part of the sky. When the solar zenith angle decreases (line 2), the path through ozone layer gets shorter and the sky color should turn redder, especially far from the dusk segment. If scattering on the aerosol particles appears, it will cause additional red excess of brightness in the dusk/dawn area by the reason described above. Change of sky color in the different points of solar vertical during the twilight is used to detect the stratospheric aerosol scattering and to explore its properties.

## 3. Aerosol scattering analysis

As it was shown in lidar (Burlakov et al., 2011) and space limb (Bourassa et al., 2008) measurements, aerosol density above the Junge layer decreases with altitude and becomes small in upper stratosphere. The brightness excess in dusk area shown in Figure 1 fades at solar zenith angle about 95-96°, that corresponds to effective scattering altitude about 35-40 km for $\zeta = +45°$.

We introduce the observed color index of the sky as the value

$$C_{RB}(\zeta, z) = \ln \frac{I_R(\zeta, z)}{I_B(\zeta, z)}. \qquad (1)$$

Here $I$ is the background intensity, $\zeta$ is the sky point position in the solar vertical, $z$ is the solar zenith angle, R and B are the color channels. Figure 3 shows the dependencies of color indexes $C_{RB}$ on solar zenith angle for different $\zeta$ values from $-45°$ to $+45°$, the evening twilight of March, 27, 2016. As it was expected, this value increases (the color turns redder) during the light stage of twilight compared with dark stage. The trend gets faster from the zenith (bold line) to the dusk-opposite region (dashed lines), and much faster from the zenith to the dusk area (solid lines), that can be related with Chappuis absorption with multiple scattering and stratospheric aerosol, respectively.

We see that color variations along the solar vertical are minimal at solar zenith angle $z_0$ equal to 96°. This time the trends are practically the same for all $\zeta$ values. This is logical since the single effective scattering takes place above the ozone and stratospheric aerosol layer. We take this moment as the reference and check the color index evolution to the lighter stage of twilight, introducing the value:

$$D_{RB}(\zeta, z) = C_{RB}(\zeta, z) - C_{RB}(\zeta, z_0). \qquad (2)$$

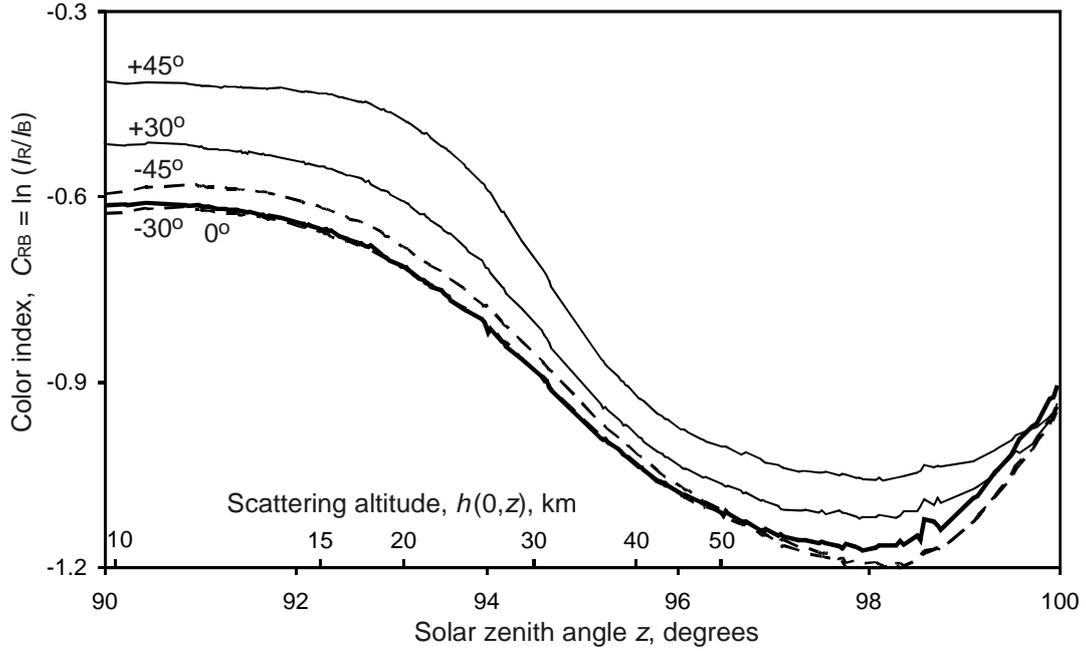

*Figure 3. Color index of the sky background in solar vertical points during the evening twilight of March, 27, 2016, the values of ζ are denoted near the curves.*

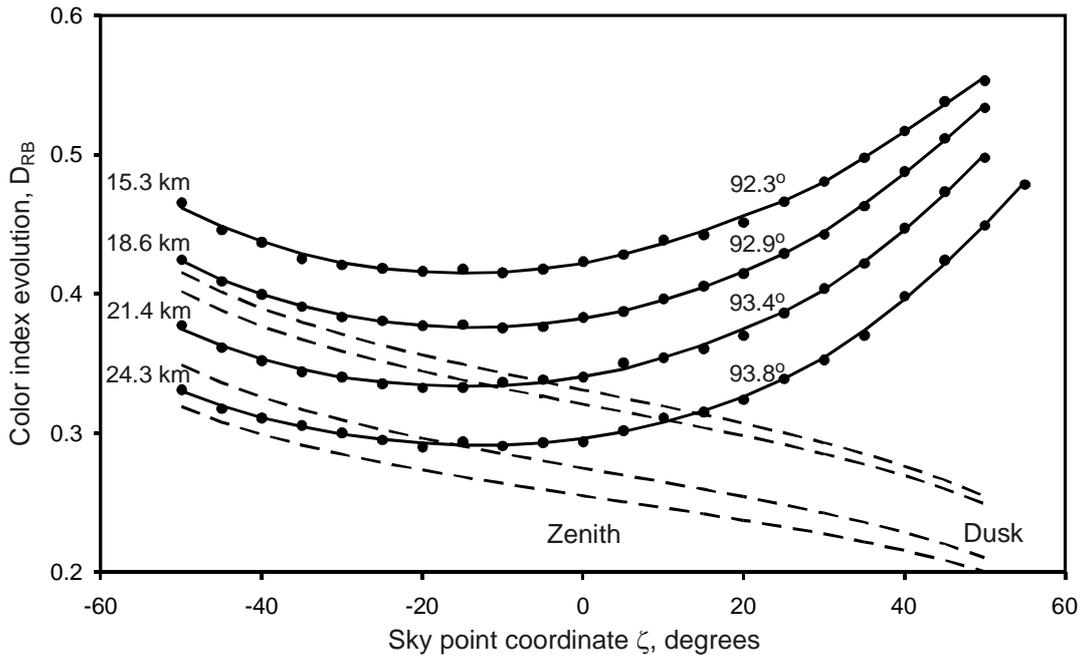

*Figure 4. Difference of sky color indexes at zenith angles z (denoted near the curves) and 96°, the same twilight as in Fig.3. Solid lines correspond to best-fit model of stratospheric aerosol, dashed lines refer to aerosol-free case. The values of effective scattering altitudes at the zenith are denoted in the left.*

The examples of the dependences $D_{RB}$ (ζ) for different solar zenith angles (the same twilight) are shown in Figure 4. It shows a fast increase at large positive ζ on a background of gradual decrease (dashed lines), as it was expected.

Building the numerical scheme of aerosol scattering separation, we assume it to be negligibly small in B channel. We can do it basing on effects in Figure 1, this assumption can lead just to underestimation of aerosol scattering contribution in R channel. Physically it is defined not only by significant decrease of aerosol to Rayleigh scattering ratio in B channel but also the difference of effective scattering altitudes in these two channels. The brightness of the sky background measured in R channel is

$$I_R(\zeta, z) = I_{0R}(\zeta, z) + I_{AR}(\zeta, z). \tag{3}$$

Here $I_A$ is the stratospheric aerosol scattering intensity, $I_0$ is the background in the case of clear stratosphere. If $I_A$ is a small admixture to $I_0$, then we can write the equation for the sky color index:

$$C_{RB}(\zeta, z) = \ln \frac{I_{0R}(\zeta, z) + I_{AR}(\zeta, z)}{I_B(\zeta, z)} = \ln \frac{I_{0R}(\zeta, z)}{I_B(\zeta, z)} + \ln \frac{I_{0R}(\zeta, z) + I_{AR}(\zeta, z)}{I_{0R}(\zeta, z)} \approx$$
$$\approx \ln \frac{I_{0R}(\zeta, z)}{I_B(\zeta, z)} + \frac{I_{AR}(\zeta, z)}{I_{0R}(\zeta, z) + (I_{AR}(\zeta, z)/2)}. \tag{4}$$

The last term is good approximation of logarithm of total and aerosol-free brightness values in R channel. The denominator of the term is the mean quantity of these two values can be also written as $(I_R(\zeta, z) - I_{AR}(\zeta, z)/2)$, we denote it as $I_{MR}(\zeta, z)$. Assuming that aerosol effects are negligibly small at deeper twilight ($z_0 = 96°$), we write the equation for color index evolution:

$$D_{RB}(\zeta, z) = \ln \frac{I_{0R}(\zeta, z)}{I_B(\zeta, z)} + \frac{I_{AR}(\zeta, z)}{I_{MR}(\zeta, z)} - \ln \frac{I_{0R}(\zeta, z_0)}{I_B(\zeta, z_0)} = D_{0RB}(\zeta, z) + \frac{I_{AR}(\zeta, z)}{I_{MR}(\zeta, z)}. \tag{5}$$

The term $D_0$ is related with the color change due to Chappuis absorption of atmospheric ozone and effects of multiple scattering. We assume it to be linear by the length of emission path through the stratosphere before scattering (see Figure 2):

$$D_{0RB}(\zeta, z) = A_0 + const \cdot d = A_0 + A_1 tg(z - \zeta - 90^O). \tag{6}$$

The aerosol scattering brightness far from horizon is being found in a form:

$$I_{AR}(\zeta, z) = \frac{F(r, \sigma, z - \zeta) P(h(\zeta, z))}{\cos \zeta} e^{-\frac{E_R}{\cos \zeta}}. \tag{7}$$

Here $F$ is the first component of Mie scattering matrix, refraction index for sulfate particles is taken to be equal to 1.43, $r$ and $\sigma$ are the parameters of log-normal particle size distribution ($r$ is the mean radius, and $\sigma$ is the exponent of standard deviation of radius logarithm). $E_R$ is the vertical extinction value for R band. The scattering angle is equal to $(z - \zeta)$, disregarding the refraction (about 0.2° for effective path).

Function $P$ is related with aerosol vertical profile and characterizes the dependence of aerosol scattering brightness on the effective altitude of scattering $h$. We take $h$ value as corresponding to the solar emission path to the scattering point with optical depth $\tau = 1$. It is the altitude of most effective scattering (the atmospheric density decreases upwards, and extinction of solar emission gets stronger downwards). The value of $h$ is calculated using the atmospheric model with real temperature and ozone vertical profiles for each observation date by EOS Aura/MLS data (EOS Team, 2011ab). These values for zenith ($\zeta = 0$) are denoted on the *x*-axis in Figures 1 and 3. If we move to another sky point, the altitude $h$ will change. For every fixed value of solar zenith angle $z$ and corresponding short interval of $h$ we assume the dependence $P(h)$ to be exponential:

$$P(h(\zeta, z)) = P_0(h(0, z)) e^{-K(h(\zeta, z) - h(0, z))}. \tag{8}$$

During the light twilight conditions, the difference of effective scattering altitudes in the solar vertical point (ζ) and zenith, $(h(\zeta,z)-h(0,z))$, does not exceed 2-3 km, being negative in the dusk/dawn area and positive in the opposite sky part. Substituting (6-8) into (5), we have

$$D_{RB}(\zeta,z) = A_0 + A_1 tg(z-\zeta-90^o) + \frac{F(r,\sigma,z-\zeta)\,P_0(h(0,z))e^{-K(h(\zeta,z)-h(0,z))}}{I_{MR}(\zeta,z)\cdot\cos\zeta}e^{-\frac{E_R}{\cos\zeta}}. \quad (9)$$

This equation has the unknown parameters: $A_0$, $A_1$, $P_0$, $r$, and $\sigma$. We have a number of measurements for different ζ values (the step is 5°). However, this system is hard to solve directly, since the different $(r-\sigma)$ pairs can correspond to very close scattering functions $F$ (the wider is distribution, or the more σ, the less is mean radius $r$). This effect is well-known, and different methods of determination of particle size distribution lead to the results as the lines in $(r-\sigma)$ diagram (Bourassa et al., 2008) or mean particle radius $r$ for fixed value of σ. The same is true for noctilucent clouds (Ugolnikov et al., 2016). We also don't initially know the amount of aerosol altitude gradient $K$ and the value $I_M$ in R channel.

To solve this problem, we use iteration method. For the first approximation, we assume the value $K$ to be same as for Rayleigh scattering and $I_{MR}=I_0=I_R$ (the contribution of aerosol scattering is small). Then we find the particle radius value $r$ for the lognormal distribution with σ=1.6 (according to (Deshler et al., 2003)). The system becomes non-linear by one unknown parameter, $r$, and linear by three other ones: $A_0$, $A_1$, and $P_0$, being quite easy to solve by least squares method. Completing this procedure for different solar zenith angles $z$, we find the intensity of stratospheric aerosol scattering $I_{AR}(\zeta,z)$. This allows finding the value of $K$ for definite altitude $h$:

$$K(h) = -\frac{dP_0(h(0,z))}{P_0(h(0,z))dh}. \quad (10)$$

We can also find the quantity

$$I_{MR}(\zeta,z) = I_R(\zeta,z) - (I_{AR}(\zeta,z)/2) \quad (11)$$

and use it together with $K(h)$ at the next iteration step in equation (9). Owing to low contribution of stratospheric aerosol scattering in the total twilight background and small variations of the altitude $h$ along the solar vertical (equation (8)), the process reaches the result fast, and we find the aerosol scattering field $I_{AR}(\zeta,z)$. At the last iteration step, the solution is being found for different σ values with step equal to 0.1. The best-fit models are shown by solid lines in the Figure 4 for the same $z$ as observational dots, corresponding effective scattering altitudes $h(0,z)$ for the zenith are also noted. Typical best-fit parameters for March, 27 are ($r$=0.07 microns, σ=1.6), ($r$=0.09 microns, σ=1.5), or ($r$=0.12 microns, σ=1.4).

## 4. Particle size distributions

Results of size distribution retrieval on $(r-\sigma)$ diagram for $z$=93.4°, $h(0,z)$=21.4 km are shown in Figure 5. The best-fit solution for this moment corresponds to model with $r$=0.091 microns and σ=1.5 (dot in the figure), possible solution area is extended to (smaller mean radii – wider distributions) and (larger mean radii – narrower distributions). The areas for single, double and triple error are shown. Dashed line shows the example of best-fit result of OSIRIS limb scattering spectroscopy (Bourassa et al., 2008), that is in good agreement with twilight data.

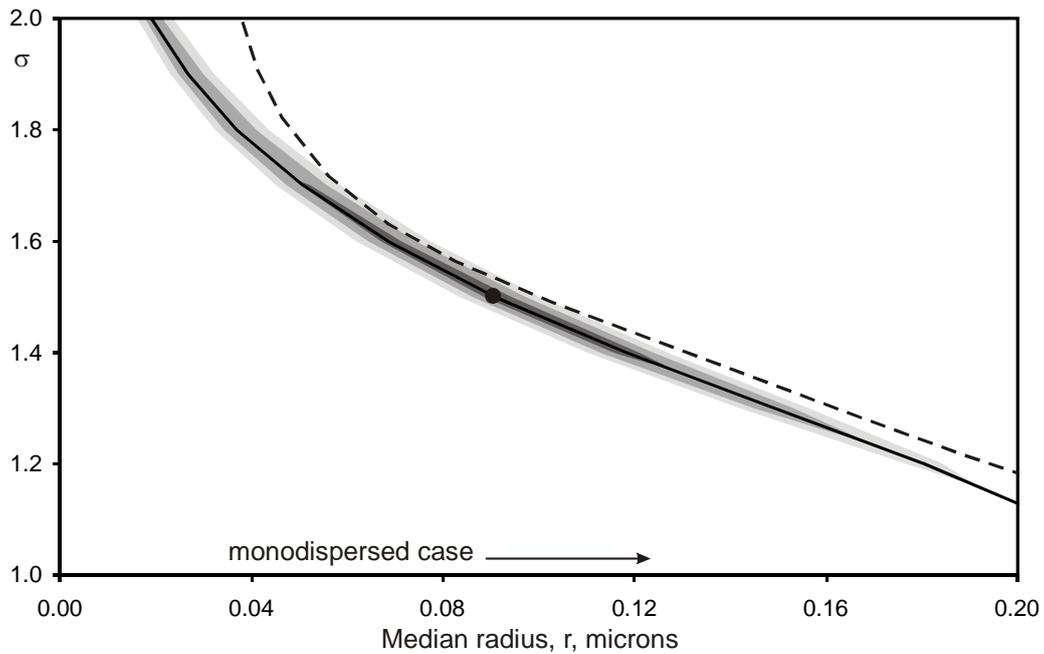

*Figure 5. Retrieved characteristics of particle log-normal size distribution: solid line and gray areas (single, double and triple error) − this work, 21.4 km, dashed line − Bourassa et al. (2008).*

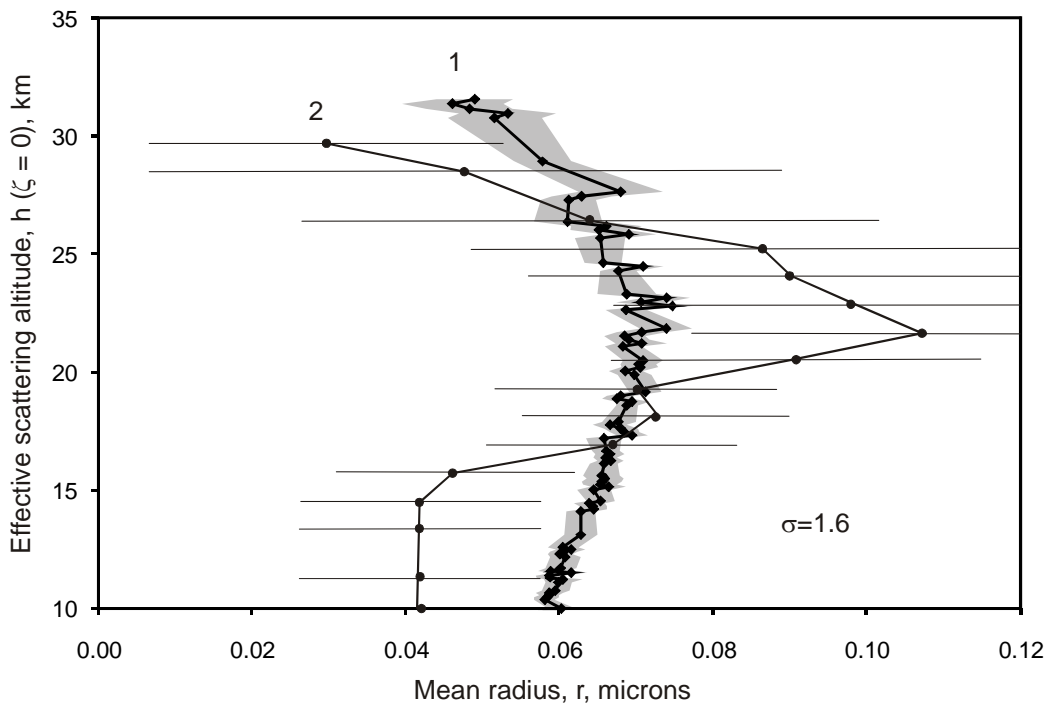

*Figure 6. Vertical profiles of mean particle radius, 1 − this work, the same twilight as in Fig. 3-5, 2 − Bourassa et al. (2008).*

Figure 6 shows the vertical dependency of mean particle radius for lognormal distribution with $\sigma = 1.6$ compared with profile obtained in (Bourassa et al., 2008) for the same $\sigma$. Both profiles show the maximum of particle size near 22 km. However, this maximum by the twilight data is less remarkable. This blurring effect can be explained by the thickness of "twilight layer", a wide range of altitudes making the contribution to the aerosol scattering during the definite twilight stage.

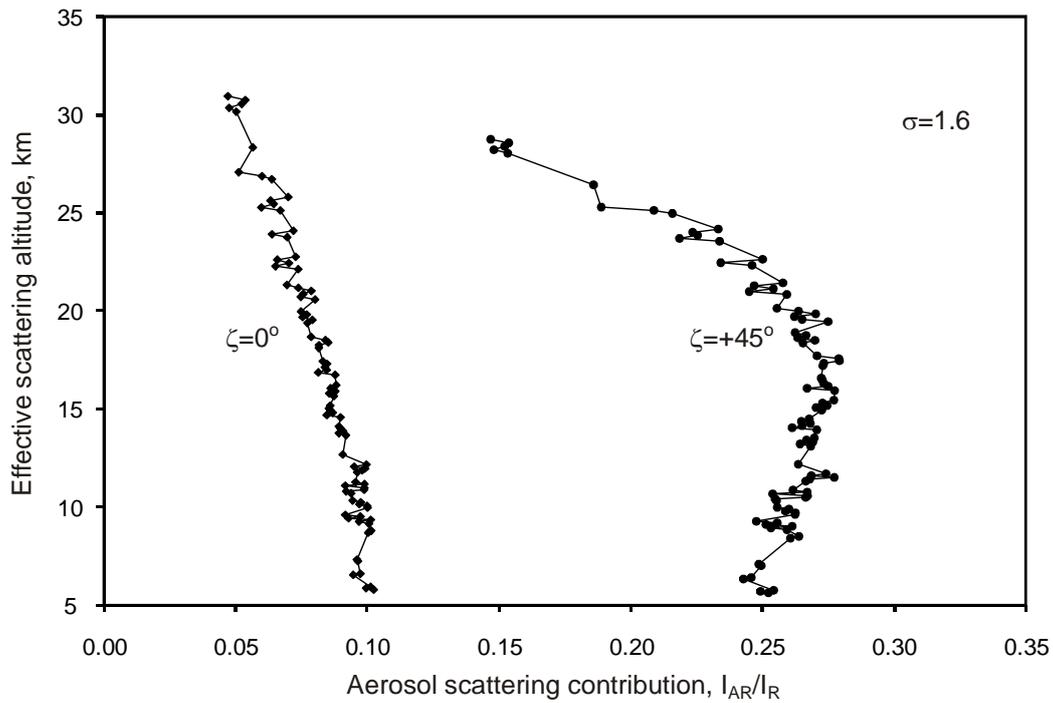

*Figure 7. Altitude profile of aerosol scattering contribution in the sky background for different solar vertical points, the same twilight as in Fig. 3-5.*

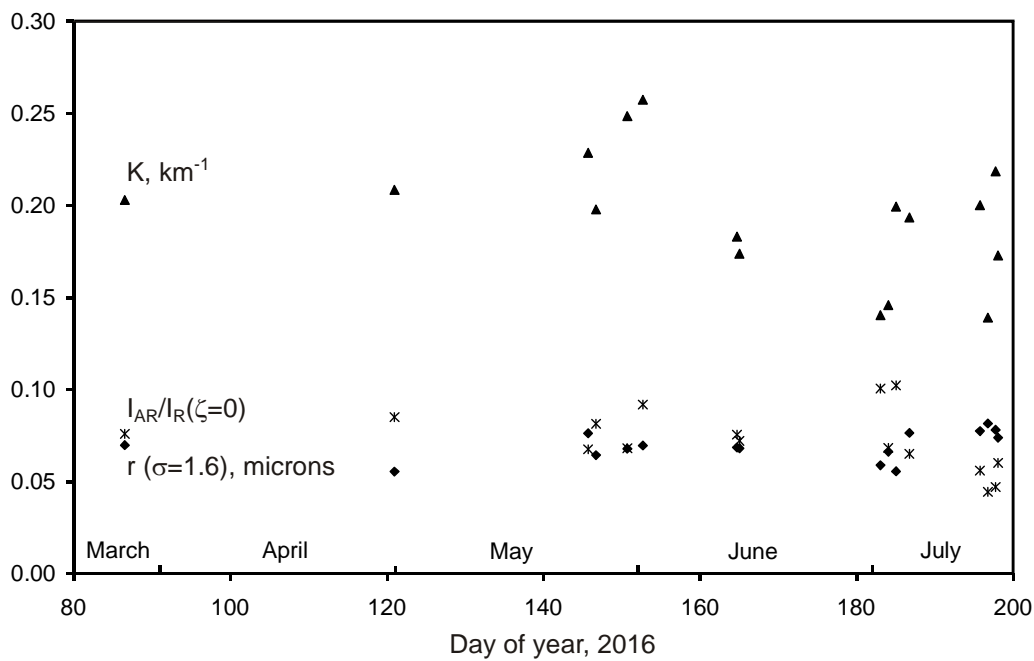

*Figure 8. Evolution of contribution to the sky brightness in the zenith, mean particle radius and altitude gradient of aerosol at 20 km in spring and summer, 2016.*

The dependencies of aerosol scattering contribution in total sky brightness on a effective scattering altitude are shown in Figure 7 for two sky positions (zenith, $\zeta=0°$, and dusk area, $\zeta=+45°$). It should be noted that for any fixed moment the effective scattering altitude is not the same in these points. The value shown in figure is not equal to the ratio of aerosol and Rayleigh scattering intensity, since the sky background is also contributed by multiple scattering with intensity about 30% of the total value (Ugolnikov and Maslov, 2002). This effect together with ozone Chappuis absorption and difference of effective thickness of Rayleigh and aerosol scattering layers makes difficult to estimate the total extinction of stratospheric aerosol. Background aerosol scattering is significant only in dusk/dawn area. For all other sky areas its contribution is just about several percents, that can be hard to detect.

The results shown in Figures 3-7 refer to the evening twilight of March, 27, 2016, the first twilight of observations with all-sky camera and color CCD. Figure 8 shows the temporal evolution of background aerosol characteristics during the observational period of spring and summer, 2016 for effective scattering altitude 20 km: the contribution of aerosol scattering in the total background in the zenith $I_{AR}/I_R$, particle radius $r$ (lognormal distribution with $\sigma = 1.6$), and altitude gradient of vertical aerosol distribution $K$. We don't see any significant seasonal trend of stratospheric aerosol characteristics during the spring and summer months of 2016.

## 5. Discussion and conclusion

In this paper we consider the effect of light scattering on stratospheric aerosol particles that can be observed by multi-wavelength observations during the twilight. This effect is quite significant even in the case of background stratospheric aerosol, it can significantly increase after volcanic eruptions. Fortunately, this observational effect has strong wavelength dependence, appearing in red spectral range. Use of RGB CCD-cameras allows to investigate it numerically, comparing the sky background properties in R and B channels. The method described here does not require polarization measurements of the sky background. However, polarization data together with extinction values measured by satellites can significantly increase the accuracy of size distribution of particles.

The contribution of aerosol scattering in the total sky background reaches about 25% in the dusk/dawn area. In the opposite sky part this value is not more than several percents. The backscattering ratio is expected to be even less, that makes this type of aerosol more difficult to investigate by lidar technique.

The results of size distribution and its vertical profile are in good agreement with other methods of stratosphere aerosol sounding. Typical mean radius of particles is about 0.07-0.08 microns for $\sigma = 1.6$, the size and contribution of this fraction of sky background reaches maximum at effective scattering altitude about 22 km.

The method described here can be the basis of systematical monitoring of stratospheric aerosol by a large number of color all-sky cameras installed in northern latitudes for aurora and noctilucent clouds observation. This will help to fix possible trends of aerosol characteristics and effects of volcanoes eruptions during upcoming years.

**Acknowledgments**

Authors are grateful to Andrey M. Tatarnikov (Sternberg Astronomical Institute, Moscow State University) for his help in observation preparations. The work is supported by Russian Foundation for Basic Research, grant No.16-05-00170-a.